\documentclass[twoside,twocolumn,9pt]{article}
\usepackage{extsizes}
\usepackage[super,sort&compress,comma]{natbib} 
\usepackage[version=3]{mhchem}
\usepackage[left=1.5cm, right=1.5cm, top=1.785cm, bottom=2.0cm]{geometry}
\usepackage{balance}
\usepackage{mathptmx}
\usepackage{sectsty}
\usepackage{graphicx} 
\usepackage{lastpage}
\usepackage[format=plain,justification=justified,singlelinecheck=false,font={stretch=1.125,small,sf},labelfont=bf,labelsep=space]{caption}
\usepackage{float}
\usepackage{fancyhdr}
\usepackage{fnpos}
\usepackage[english]{babel}
\addto{\captionsenglish}{%
	
}
\usepackage{array}
\usepackage{droidsans}
\usepackage{charter}
\usepackage[T1]{fontenc}
\usepackage[usenames,dvipsnames]{xcolor}
\usepackage{setspace}
\usepackage[compact]{titlesec}
\usepackage{hyperref}
\usepackage{chemformula}
\usepackage{booktabs, multirow}
\usepackage{soul}

\usepackage{epstopdf}

\usepackage[columnwise]{lineno}

\definecolor{cream}{RGB}{222,217,201}

\begin{document}
	
	\pagestyle{fancy}
	\thispagestyle{plain}
	\fancypagestyle{plain}{
		\renewcommand{\headrulewidth}{0pt}
	}
	
	\makeFNbottom
	\makeatletter
	\renewcommand\LARGE{\@setfontsize\LARGE{15pt}{17}}
	\renewcommand\Large{\@setfontsize\Large{12pt}{14}}
	\renewcommand\large{\@setfontsize\large{10pt}{12}}
	\renewcommand\footnotesize{\@setfontsize\footnotesize{7pt}{10}}
	\makeatother
	
	\renewcommand{\thefootnote}{\fnsymbol{footnote}}
	\renewcommand\footnoterule{\vspace*{1pt}%
		\color{cream}\hrule width 3.5in height 0.4pt \color{black}\vspace*{5pt}} 
	\setcounter{secnumdepth}{5}
	
	\makeatletter 
	\renewcommand\@biblabel[1]{#1}            
	\renewcommand\@makefntext[1]%
	{\noindent\makebox[0pt][r]{\@thefnmark\,}#1}
	\makeatother 
	\renewcommand{\figurename}{\small{Fig.}~}
	\sectionfont{\sffamily\Large}
	\subsectionfont{\normalsize}
	\subsubsectionfont{\bf}
	\setstretch{1.125} 
	\setlength{\skip\footins}{0.8cm}
	\setlength{\footnotesep}{0.25cm}
	\setlength{\jot}{10pt}
	\titlespacing*{\section}{0pt}{4pt}{4pt}
	\titlespacing*{\subsection}{0pt}{15pt}{1pt}
	
	\fancyfoot{}
	\fancyfoot[LO,RE]{\vspace{-7.1pt}\includegraphics[height=9pt]{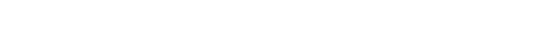}}
	\fancyfoot[CO]{\vspace{-7.1pt}\hspace{11.9cm}\includegraphics{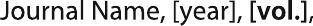}}
	\fancyfoot[CE]{\vspace{-7.2pt}\hspace{-13.2cm}\includegraphics{head_foot/RF}}
	\fancyfoot[RO]{\footnotesize{\sffamily{1--\pageref{LastPage} ~\textbar  \hspace{2pt}\thepage}}}
	\fancyfoot[LE]{\footnotesize{\sffamily{\thepage~\textbar\hspace{4.65cm} 1--\pageref{LastPage}}}}
	\fancyhead{}
	\renewcommand{\headrulewidth}{0pt} 
	\renewcommand{\footrulewidth}{0pt}
	\setlength{\arrayrulewidth}{1pt}
	\setlength{\columnsep}{6.5mm}
	\setlength\bibsep{1pt}
	
	\makeatletter 
	\newlength{\figrulesep} 
	\setlength{\figrulesep}{0.5\textfloatsep} 
	
	\newcommand{\topfigrule}{\vspace*{-1pt}%
		\noindent{\color{cream}\rule[-\figrulesep]{\columnwidth}{1.5pt}} }
	
	\newcommand{\botfigrule}{\vspace*{-2pt}%
		\noindent{\color{cream}\rule[\figrulesep]{\columnwidth}{1.5pt}} }
	
	\newcommand{\dblfigrule}{\vspace*{-1pt}%
		\noindent{\color{cream}\rule[-\figrulesep]{\textwidth}{1.5pt}} }
	
	\makeatother
	
	\twocolumn[
	\begin{@twocolumnfalse}
		{\includegraphics[height=30pt]{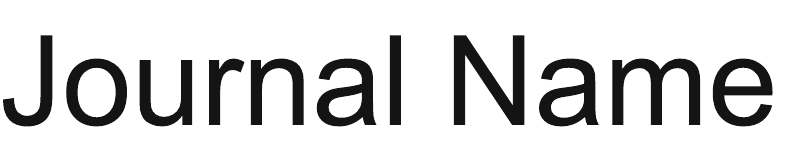}\hfill\raisebox{0pt}[0pt][0pt]{\includegraphics[height=55pt]{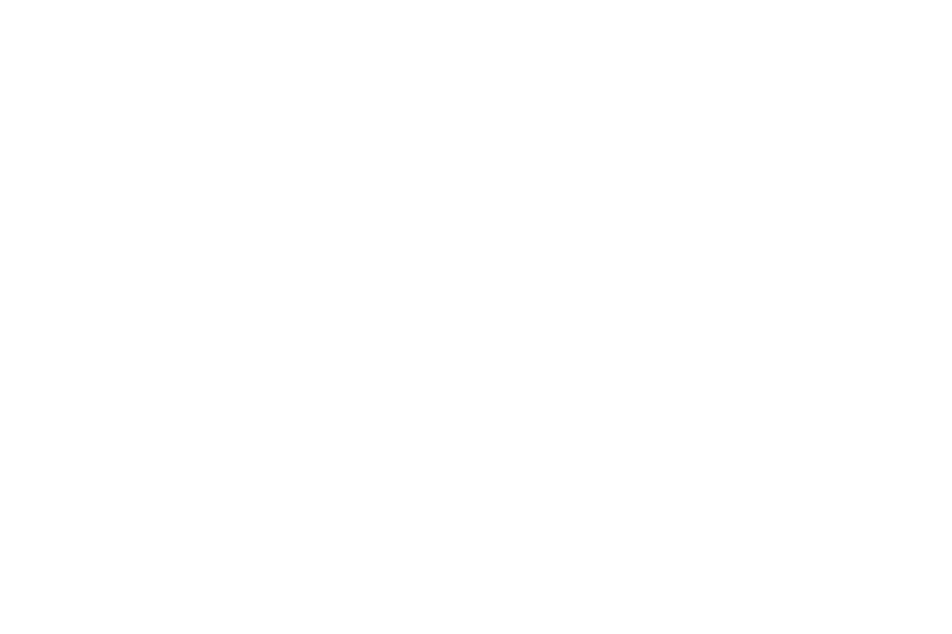}}\\[1ex]
			\includegraphics[width=18.5cm]{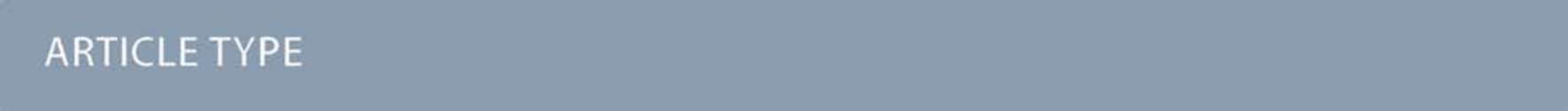}}\par
		\vspace{1em}
		\sffamily
		\begin{tabular}{m{4.5cm} p{13.5cm} }
			
			\includegraphics{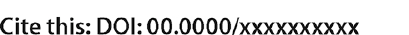} & \noindent\LARGE{\textbf{Coincident angle-resolved state-selective photoelectron spectroscopy of acetylene molecules: a candidate system for time-resolved dynamics}} \\
			\vspace{0.3cm} & \vspace{0.3cm} \\
			
			& \noindent\large{S. Mandal\textit{$^{a}$}, R. Gopal\textit{$^{b}$}, H. Srinivas\textit{$^{c}$}, A. D’Elia\textit{$^{d}$}, A. Sen\textit{$^{a}$}, S Sen\textit{$^{e}$}, R. Richter\textit{$^{f}$}, M. Coreno\textit{$^{g,h}$}, B. Bapat\textit{$^{a}$}, M. Mudrich\textit{$^{i,j}$}, V. Sharma\textit{$^{e\dag}$}} and 
			S. R. Krishnan\textit{$^{j\ast}$} \\
			
			\includegraphics{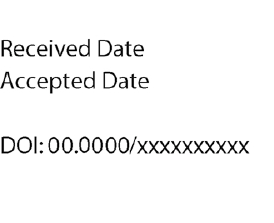} & \noindent\normalsize{
			The acetylene-vinylidene system serves as a benchmark for investigations of ultrafast dynamical processes where the coupling of the electronic and nuclear degrees of freedom provides a fertile playground to explore the femto- and sub-femto-second physics with coherent extreme-ultraviolet (EUV) photon sources both on the  table-top as well as  free-electron lasers. We focus on detailed investigations of this molecular system in the photon energy range $19...40$ \textrm{eV} where EUV pulses can probe the dynamics effectively. We employ photoelectron-photoion coincidence (PEPICO) spectroscopy to uncover hitherto unrevealed aspects of this system. In this work, the role of excited states of the \ce{C2H2+} cation, the primary photoion, is specifically addressed. From photoelectron energy spectra and angular distributions, the nature of the dissociation and isomerization channels is discerned. Exploiting the $4\pi$-collection geometry of velocity map imaging spectrometer, we not only probe pathways where the efficiency of photoionization is inherently high but also perform PEPICO spectroscopy on relatively weak channels.}
		\end{tabular}
		
	\end{@twocolumnfalse} \vspace{0.6cm}
	
	]
	
	\renewcommand*\rmdefault{bch}\normalfont\upshape
	\rmfamily
	\section*{}
	\vspace{-1cm}

	
	\footnotetext{\textit{$^{a}$Indian Institute of Science Education and Research, Pune 411008, India}}
	\footnotetext{\textit{$^{b}$Tata Institute of Fundamental Research, Hyderabad 500107, India}}
	\footnotetext{\textit{$^{c}$Max-Planck-Institut f\"{u}r Kernphysik, 69117 Heidelberg, Germany}}
	\footnotetext{\textit{$^{d}$IOM-CNR, Laboratorio TASC, Basovizza SS-14, km 163.5, 34149 Trieste, Italy}}
	\footnotetext{\textit{$^{e}$Indian Institute of Technology Hyderabad, Kandi 502285, India; E-mail: vsharma@phy.iith.ac.in}}
	\footnotetext{\textit{$^{f}$Elettra-Sincrotrone Trieste, 34149 Basovizza, Italy}}
	\footnotetext{\textit{$^{g}$Istituto di Struttura della Materia - Consiglio Nazionale delle Ricerche, 34149 Trieste, Italy}}
	\footnotetext{\textit{$^{h}$INFN – LNF, via Enrico Fermi 54, 00044 Frascati, Italy}}
	\footnotetext{\textit{$^{i}$Department of Physics and Astronomy, Aarhus University, 8000 Aarhus C, Denmark}}
	\footnotetext{\textit{$^{j}$Department of Physics and QuCenDiEm-Group, Indian Institute of Technology Madras, Chennai 600036, India; E-mail: srkrishnan@iitm.ac.in}}
	
	
	
	
	\section{Introduction}
	One of the outstanding problems of interest in time-resolved spectroscopy and quantum dynamics of molecular systems is phenomena involving the interplay between nuclear motion and electron dynamics\cite{Stolow_AnnRevPhysChem_2003, Stolow_ChemRev_2004}.  In femto- and sub-femto-second timescales, a deep understanding of these scenarios is intimately related to realizing the grand challenge of making molecular movies; "watching" chemical reactions take place\cite{Zewail_JPCA_2000}. Among important aspects of the physics of systems beyond the Born-Oppenheimer approximation\cite{Cederbaum_JCP_2008}, decoupling nuclear and electronic dynamics, the role of conical intersections\cite{Worth_AnnRev_2004}, shape resonances\cite{Nandi_SciAdv_2000}, and fast rearrangements within molecules\cite{jiang_PRL_2010, Liekhus-Schmaltz_NatCommn_2015, Galbraith_PCCP_2017} are of particular interest. Proton migration ensuing in the rearrangement of photoexcited molecular systems has a prominent place not only owing to the intriguing physics, but also due to its importance in biological systems; this plays a key role in processes underlying human vision\cite{Schoenlein_Science_1991}, photosynthesis\cite{Kreutz_1970}, proton tunneling in DNA\cite{Lowdin_RevModPhys_1963} and radiation damage\cite{Rein_Science_1964}, to name a few. 
	
	The acetylene-vinylidene system has long served as the benchmark for investigations of isomerization especially on ultrafast timescales \cite{jiang_PRL_2010, Burger_SD_2018, Li_NatCommn_2017} as well as in static  spectroscopy and theoretical investigations\cite{Boye_JCP_2006, Zyubina_JCP_2005, Alagia_JCP_2012, Krishnan_CPL_1981}. Both the photoexcitation of outer-valence electrons\cite{jiang_PRL_2010} in the extreme-ultraviolet as well as core-shell electrons in the hard-xray regimes can effect isomerization\cite{Liekhus-Schmaltz_NatCommn_2015}. Understanding this system paves way for investigating the dynamics of proton migration in larger systems such as benzene\cite{Galbraith_PCCP_2017}, and proton conduction in covalently bonded molecules\cite{Vilciauskas_NatChem_2012} and weakly bound aggregates such as  bio-interfaces\cite{Branden_PNAS_2006}.	In order to perform time-resolved spectroscopy of the acetylene-vinylidene system, an intimate knowledge of not only the neutral molecule but also the residual ion and more importantly, the details of photoelectron energies and angular distributions is essential: For example, transient absorption, laser-induced fluorescence or resonant multiphoton ionization methods which are popular in this context are effective when the system is spectroscopically well-characterized. The dynamics of wavepacket resulting from the finite bandwidth of interrogating pulses can be traced effectively when the states involved are known a priori.
	
	In this article, we use photoelectron imaging in coincidence with photoion spectrometry to uncover the details of this benchmark system in the spirit of preparing the ground for further investigations of this system using table-top as well as free-electron pulsed laser sources. While, reports on the transient dynamics of this system have been published, our recent investigations of this molecular system embedded in \ce{He} nanodroplet-environment motivate further time-resolved studies\cite{Mandal_PCCP_2020} to estimate the time scale of environment assisted Penning ionization of \ce{C2H2} from higher lying states of \ce{He}$^\ast$ ($n=4$) band. One of the key advantages of photoelectron spectroscopy is that it can be readily applied in time-resolved studies bringing with it the advantage of accessing the entire reaction coordinate even when the electronic and vibrational states evolve in time. Thus, photon energy dependent study of partial cross sections and photoelectron angular distributions has proved to be a useful tool to probe different resonant autoionization processes and shape resonance phenomena in molecular species\cite{Parr_JCP_1982, Parr_PRL_1981, Codling_JPhysB_1981, West_JPhysB_1981}. 
	
	In these investigations of acetylene (\ce{C2H2}) photoionization by the  photoelectron-photoion coincidence technique, we report the photoelectron energy spectra (PES) corresponding to different ionization channels of \ce{C2H2+} along with their accompanying photoelectron angular distributions (PADs). This allows us to discern PES and PADs for each \ce{C2H2} fragment ion as a function of photon energy; this includes the primary photoion \ce{C2H2+} as well as those resulting from further dissociation and isomerization. The choice of the velocity-map-imaging scheme for  photoelectrons is deliberate. This technique is a preferred method for studying molecules and clusters with extreme-ultraviolet  pulses and high-harmonic generation methodology\cite{Kreibich_PRL_2001} owing to the inherently high-collection efficiency over the entire solid angle\cite{Mauritsson_PRL_2010, Kelkensberg_PRL_2011, Sansone_Nature_2010}. Thus, our results can be immediately carried forward and applied to these scenarios. 

	The key findings of this work are as follows: Firstly, we precisely characterize all the fragmentation channels and determine the electronic states responsible for producing each of these fragments. These are validated by the fact that the contributions of the highest occupied molecular orbital (HOMO) to the photoelectron spectra are in good agreement with prior theoretical work\cite{Fronzoni_CPL_2004}. However, the photoelectron angular distribution measurements do not always agree with reported theory; nor do they evidence autoionizing resonances when correlated to particular ionic fragments. But it is noteworthy that earlier computations do not match unanimously, either\cite{Fronzoni_CPL_2004, Wells_JCP_1999a}. Thus, our work provides pertinent inputs for revisions over and above the existing work. Owing to the merits of the experimental technique, we employ, we could ascertain that the less abundant ionic fragments which result from single ionization including, \ce{C2+}, \ce{CH2+}, \ce{CH^+} and \ce{C+} arise from the higher-excited ionic states. The hallmark of this article is that, to the best of our knowledge, we have for the first time measured state-selective branching ratios, photoelectron angular distributions and asymmetry parameters, as a function of photon energy, for all the relevant cationic states of primary photoion, \ce{C2H2+}, for different photoionization pathways both below and above the double ionization energy of this paradigmatic molecular system.
	
\section{Experimental methods}
	
\begin{figure}[t]
		\centering
		\includegraphics[width=0.45\textwidth]{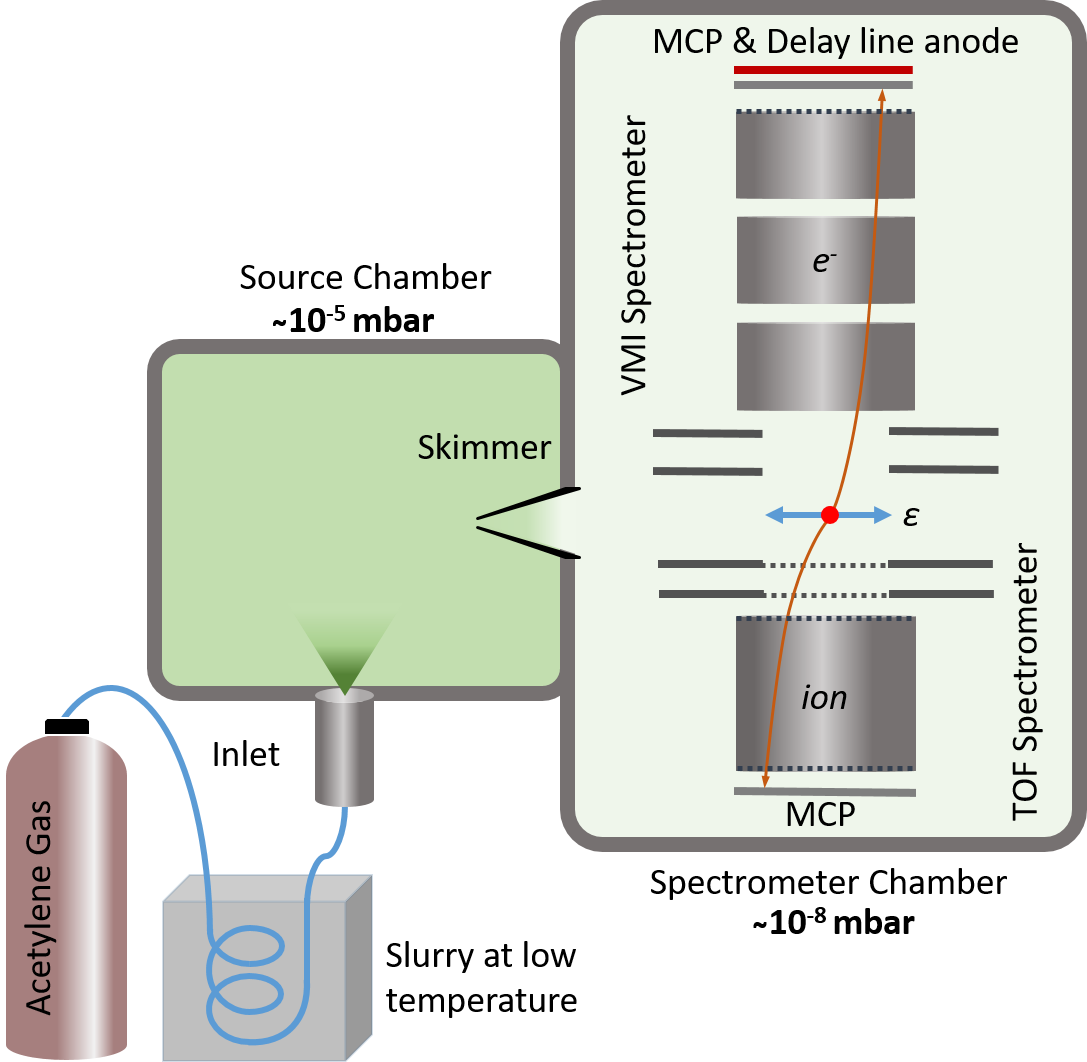}
		\caption{Schematic of the experimental setup: Acetylene gas is effused through a leak valve in the source chamber adjacent to the spectrometer chamber which is connected by a conical skimmer. Subsequently, through the skimmer, \ce{C2H2} is flooded into the spectrometer chamber where the electron-VMI and the ion-ToF spectrometers are situated. In the spectrometer chamber, \ce{C2H2} is ionized by linearly polarized EUV photon beam.}
		\label{Fig1_Ac}
	\end{figure}

	The experiments reported here were carried out at the Gasphase beamline of the Elettra Sincrotrone, Trieste. Fig.\ref{Fig1_Ac} shows the schematic diagram of the experimental setup, whose details have been published earlier \cite{OKeeffe_RSI_2011}. Here, high-purity \ce{C2H2} gas was effused into the source chamber through a dosing valve. The \ce{C2H2} gas was distilled before entering this valve to remove acetone contamination. In the distillation process, the gas mixture was passed through a slurry of ethanol and liquid \ce{N2} maintained at $-100$ \textrm{$^{\circ}C$}. The source chamber is connected to spectrometer chamber through a conical skimmer which maintains a differential pressure; \ce{C2H2} gas effuses into  spectrometer chamber which is maintained at $\sim 10^{-8}$ \textrm{mbar}, while the source chamber remains at $\sim 10^{-5}$ \textrm{mbar}.
	
	The spectrometer chamber holds two co-axial spectrometers - a velocity map imaging (VMI) spectrometer and a time-of-flight (ToF) spectrometer (cf. fig.\ref{Fig1_Ac}). A focused beam of linearly polarized extreme-ultraviolet (EUV) photons passes through the geometric centre of the two spectrometers at right angle to the spectrometer axis which is also perpendicular to its polarization axis ($\varepsilon$). Photon energies in the range between $19$ and $40$ \textrm{eV} were used in our study. We exploited the excellent photon energy definition possible at this beamline quantified by the resolving power of the monochromator upstream, $\Delta E/E\leq10^{-4}$; using a set of gratings, high-quality photon beams in the energy range $10 - 900$ \textrm{eV} are accessible here. The synchrotron ring delivers the photon beam in this case in the form of $\sim150$ \textrm{ps} pulses with typical peak intensity of $\sim15$ \textrm{W/m$^{2}$} and repetition rate of $500$ \textrm{MHz}. Here, randomly oriented \ce{C2H2} molecules are photoionized by the EUV light and the resultant photoelectrons and photoions are detected in coincidence with the VMI and ToF spectrometers, respectively. The charged particle count rate was maintained at $\sim18$ \textrm{kHz} by adjusting two slits on the photon beam path. This synchronous detection scheme of photoelectrons and photoions enables us to measure the kinetic energies and angular distributions for photoelectrons correlated to different photoions formed due to \ce{C2H2} photoionization. Therefore, unlike previous studies \cite{Kreile_CPL_1980, Parr_JCP_1982, Holland_JESRP_1998}, not only do we get the photoelectron energy distributions of \ce{C2H2} photoionization, but also this provides photoelectron energy spectra and angular distributions correlated to specific photoions and photoionization channels.
	
	We implemented Abel inversions using the well-established program, MEVELER \cite{Dick_PCCP_2014}, to obtain the full 3D velocity distribution of photoelectrons from 2D projection images captured by the VMI spectrometer. We used known photoelectron energy distribution of \ce{He} at different photon energies above the atomic \ce{He} ionization energy ($E_{i}=24.58$ \textrm{eV}) to calibrate VMI spectrometer. The average energy resolution ($\Delta E/E$) achieved by the spectrometer is about $7\%$. For one-photon ionization by linearly polarized light, under dipole approximation, the differential cross section can be expressed as:
	\begin{equation}\label{PAD}
	\frac{d\sigma}{d\Omega} = \frac{\sigma_{total}}{4\pi}[1+\beta P_{2}(\cos{\theta})].
	\end{equation}
	Since, the photoelectron velocity ($\vec{v}$) has the cylindrical symmetry along the polarization axis ($\varepsilon$), the differential cross section has no azimuthal ($\phi$) dependence. $P_{2}(\cos{\theta})$ is the second order Legendre polynomial and $\theta$ is the angle between $\vec{v}$ and $\varepsilon$. The photoelectron angular distribution (PAD) is characterized by the asymmetry parameter, $\beta$.
	
\begin{figure}[t]
		\centering
		\includegraphics[width=0.5\textwidth]{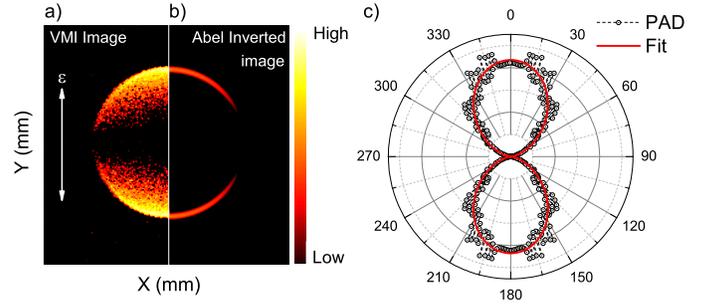}
		\caption{a) VMI distribution  and b) Abel-inverted distribution of the photoelectrons, in a logarithmic color scale, due to photoionization of effusive \ce{He} at $28$ \textrm{eV} and (c) the photoelectron angular distribution (PAD) obtained from b). The red line shows the fitting of the PAD for $\beta=2.01 \pm 0.08$, demonstrating the performance of the spectrometer.}
		\label{Fig2_Ac}
	\end{figure}

	In the current study, to obtain the value of $\beta$ specific to different ionic states, we used the following scheme: Multiple Gaussian functions are fitted to the PES to determine different ionic states and their full-width-at-half-maximum (FWHM). Then, we obtain the PAD for each state by integrating the angular photoelectron counts over the FWHM limit of each state from the Abel-inverted distribution. Finally, we fitted eq.\eqref{PAD} on the PAD to get the asymmetry parameter, $\beta$. For example, fig.\ref{Fig2_Ac} a), b) show the experimental VMI distribution and the Abel-inverted distribution of the photoelectron emitted due to photoionization of effusive \ce{He} at $28$ \textrm{eV}, respectively. Fig.\ref{Fig2_Ac} c) presents the PAD of the observed \ce{He} $1s$ ionization, where the value of $\beta$ obtained from fitting eq.\eqref{PAD} is $2.01 \pm 0.08$ which correctly correlates to the PAD of $p$- partial wave resulting from one-photon ionization \cite{OKeeffe_JOPCon_2010}.
	
\section{Results and discussion}
	
	Acetylene in its neutral ground state ($^{1}\Sigma_{g}^{+}$) has the following electronic configuration:
	$$(1\sigma_{g})^{2}(1\sigma_{u})^{2}(2\sigma_{g})^{2}(2\sigma_{u})^{2}(3\sigma_{g})^{2}(1\pi_{u})^{4}$$, 
	with $1\pi_{g}$, $3\sigma_{u}$, $4\sigma_{g}$ and $4\sigma_{u}$ being the lowest lying unoccupied orbitals. In the spectral range from $19.0$ \textrm{eV} to below double ionization energy ($E_{\textrm{di}}$ $\sim32$ \textrm{eV}), electrons are predominantly excited or ionized from the valence orbitals, $1\pi_{u}$, $3\sigma_{g}$, $2\sigma_{u}$ and $2\sigma_{g}$. Considering the  independent particle model, ionization from the $1\pi_{u}$, $3\sigma_{g}$, $2\sigma_{u}$ and $2\sigma_{g}$ orbitals leads to $X^{2}\Pi_{u}$, $A^{2}\Sigma_{g}^{+}$,  $B^{2}\Sigma_{u}^{+}$, and $C^{2}\Sigma_{g}^{+}$ states in \ce{C2H2^+}, respectively. Along with these direct ionization channels, there exists several indirect autoionizing resonances in \ce{C2H2}, where electrons are excited from the valence orbitals to the virtual orbitals upon photoabsorption. As these excitations decay to the ionic states, $X, A, B$ and $C$, the corresponding kinetic energies of photoelectrons remain same irrespective of the ionization mechanism. However, the ionization cross sections of these states and the associated photoelectron angular distributions are greatly influenced by the involved ionization processes\cite{Hayaishi_JPhysB_1982, Parr_JCP_1982, Holland_JESRP_1998, Fronzoni_CPL_2004, Wells_JCP_1999a, Langhoff_CPL_1981}. Here, we will discuss the photon energy dependent photoionization cross sections and the photoelectron angular distributions associated with different cationic states of \ce{C2H2+} both for the photoionization and for different photodissociation channels. The remainder of this article is organized as follows: First we discuss photoion mass spectra which enable us to identify distinct ionization channels characterized by the dissociation pathways of the \ce{C2H2+} ion. The mainstay of this article, photoelectron energy spectra (PES) specific to these ionization channels, as well as the photoelectron angular distributions (PADs) and the asymmetry parameters ($\beta$) particular to each ionization channel and ionic state are presented. We compare our work with existing studies wherever it is relevant to underscore new findings.  

\subsection{Photoion mass spectra and dissociation channels}

	\begin{table*}[h]\centering
		\scriptsize
		\caption{Comparison of relative ionization efficiencies ($\eta^{x}$) of the photoions ($x$) as a function of photon energy ($h\nu$) with the results obtained by \citeauthor{Hayaishi_JPhysB_1982}\cite{Hayaishi_JPhysB_1982}}
		\label{table1}
		\begin{tabular}{lrrrrrrrrrrrr}\toprule
			\multirow{3}{*}{$h\nu$ (\textrm{eV})} &\multicolumn{10}{c}{$\eta^{x}$ (arb. u.)} &\multirow{2}{*}{$\eta$ (arb. u.)} \\\cmidrule{2-11}
			&\multicolumn{2}{c}{$x=$ \ce{C2H2+}} &\multicolumn{2}{c}{$x=$ \ce{C2H+}} &\multicolumn{2}{c}{$x=$ \ce{C2+}} &\multicolumn{2}{c}{$x=$ \ce{CH2+}} &\multicolumn{2}{c}{$x=$ \ce{CH+}} & \\\cmidrule{2-12}
			&Current &Previous\cite{Hayaishi_JPhysB_1982} &Current &Previous\cite{Hayaishi_JPhysB_1982} &Current &Previous\cite{Hayaishi_JPhysB_1982} &Current &Previous\cite{Hayaishi_JPhysB_1982} &Current &Previous\cite{Hayaishi_JPhysB_1982} &Current \\\midrule
			19.0 &0.777 &0.777 &0.105 &0.112 &0.006 &0.004 &0.004 &0.002 &0.007 &0.000 &0.904 \\
			21.6 &0.739 &0.734 &0.183 &0.191 &0.014 &0.008 &0.008 &0.011 &0.029 &0.004 &0.988 \\
			23.9 &0.651 &0.628 &0.148 &0.149 &0.016 &0.024 &0.008 &0.026 &0.028 &0.019 &0.864 \\
			26.0 &0.588 &0.511 &0.137 &0.119 &0.019 &0.025 &0.007 &0.016 &0.024 &0.025 &0.784 \\
			28.0 &0.377 &0.412 &0.090 &0.087 &0.014 &0.021 &0.005 &0.012 &0.017 &0.023 &0.509 \\
			36.0 &0.283 &\textemdash &0.073 &\textemdash &0.015 &\textemdash &0.005 &\textemdash &0.027 &\textemdash &0.416 \\
			40.0 &0.209 &\textemdash &0.061 &\textemdash &0.014 &\textemdash &0.005 &\textemdash &0.029 &\textemdash &0.331 \\
			\bottomrule
		\end{tabular}
	\end{table*}

\begin{figure}[]
		\centering
		\includegraphics[width=0.5\textwidth]{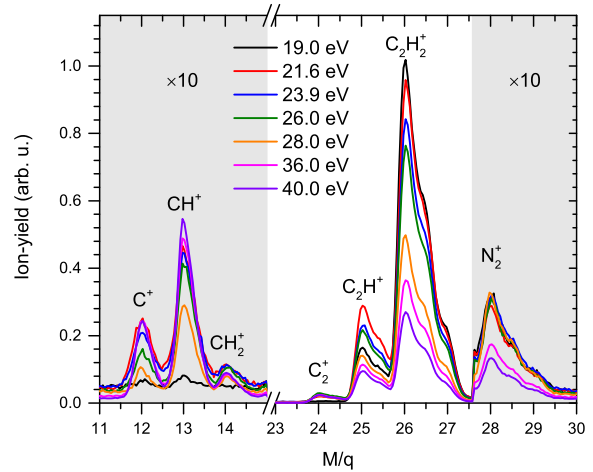}
		\caption{Photoion ToF mass spectra at different photon energies. The x-axis represents mass to charge ratio ($M/q$) of the photoions in atomic units. The shaded portion from $M/q = 11$ to $15$ and from $M/q = 27.5$ to $30$ are magnified $10$ times. Throughout, peaks are labeled by the single- and double-ionization ionic fragments from acetylene - \ce{C2H+}, \ce{C2+}, \ce{CH2+}, \ce{CH+} and \ce{C+}, along with the singly ionized parent molecular ion, \ce{C2H2+}. However, at  $h\nu = 36$ and $40$ \textrm{eV}, non-dissociative double-ionization product, \ce{C2H2^2+}, make small contribution to the peak at $M/q = 13$. While the contribution to the $M/q = 14$ peak from \ce{N+} due to residual nitrogen (\ce{N2}) ionization is tiny at lower photon energies, at $36$ and $40$ \textrm{eV} this can be significant \cite{Fennelly_ADNDT_1992}. The peak at $M/q=28$ corresponds to \ce{N2+} ions. The photoion ToF mass spectra are normalized such that the background \ce{N2+} ion yields are proportional to the partial photoionization cross-sections of \ce{N2} producing \ce{N2+} ions at the respective photon energies\cite{Fennelly_ADNDT_1992}. The relative ionization efficiencies ($\eta^{x}$) for different photoions ($x$) are directly calculated from the integral-area of the respective photoion peak at different photon energies which are proportional to the respective partial photoionization cross sections.}
		\label{Fig3_Ac}
	\end{figure}

	To identify \ce{C2H2} photoionization channels, we recorded the photoion ToF mass spectra, presented in  fig.\ref{Fig3_Ac}, at different photon energies. We observe several fragmented ions, \ce{C2H+}, \ce{C2+}, \ce{CH2+}, \ce{CH+} and \ce{C+} as well as unfragmented parent molecular ion, \ce{C2H2+}. Each of these fragmented ions represents a distinct photodissociation channel, where the respective ionic fragment is accompanied by undetected neutrals. Among these ionic products, \ce{C2H2+} and \ce{C2H+} ions are the most abundant ionic species, constituting $\sim95\%$ of total ion-yield, while the other fragments comprise the remaining fraction. Notably, in fig.\ref{Fig3_Ac} the ion-yields of the photoions ($x$), where $x$ represents \ce{C2H2+}, \ce{C2H+}, \ce{C2+}, \ce{CH2+}, \ce{CH+} and \ce{C+}, vary with photon energy, evidencing the corresponding dependence of the relative ionization efficiencies ($\eta^{x}$) of the channels involved on the same parameter.
	
	\citeauthor{Hayaishi_JPhysB_1982}\cite{Hayaishi_JPhysB_1982} extensively studied the photoionization dynamics of \ce{C2H2} by measuring the photoion-yields of \ce{C2H2+}, \ce{C2H+}, \ce{C2+}, \ce{CH2+} and \ce{CH+} as a function of photon energy. They discussed the appearance energies of these ions as well as assigned electronic excitations that result in these photoions from \textit{ab initio} theoretical calculations. Here, the relative ionization efficiencies ($\eta^{x}$) are calculated from the integral area of different photoion ($x$) peaks in the TOF mass spectra as a function of photon energy, shown in fig.\ref{Fig3_Ac}. Herein, the photoion ToF mass spectra are normalized such that the total ion-yields of background \ce{N2+} ions at different photon energies are proportional to the respective partial ionization cross section of \ce{N2+} from \ce{N2} photoionization\cite{Fennelly_ADNDT_1992}. Since, we kept the data acquisition time and the spectrometer chamber pressure at the same values for all the measurements at different photon energies, assuming the identical detection efficiencies of different photoions, the relative ionization efficiencies calculated here are proportional to the respective partial photoionization cross sections of \ce{C2H2} photoionization \cite{Cooper_CP_1988}. Table.\ref{table1} shows the comparison of the observed relative ionization efficiencies ($\eta^{x})$ of \ce{C2H2+}, \ce{C2H+}, \ce{C2+}, \ce{CH2+} and \ce{CH+} ions with the results obtained by \citeauthor{Hayaishi_JPhysB_1982}\cite{Hayaishi_JPhysB_1982}. The reported relative ionization efficiencies ($\eta^{x}$) scaled suitably so that the ionization efficiency of \ce{C2H2+} at $19.0$ \textrm{eV} photon energy, are matched with that of  \citeauthor{Hayaishi_JPhysB_1982}\cite{Hayaishi_JPhysB_1982}. In this context, it is important to note that, in the work of \citeauthor{Hayaishi_JPhysB_1982}\cite{Hayaishi_JPhysB_1982} at $h\nu=15.3$ \textrm{eV} the relative ionization efficiency of \ce{C2H2+} is equal to $1$ \textrm{arb. u.}. In table.\ref{table1}, $\eta$ represents the total relative ionization efficiency of cumulative all \ce{C2H2} photoions shown in fig.\ref{Fig3_Ac}.
	
	It is encouraging to note that there is a good agreement between our results and corresponding values from earlier  studies for \ce{C2H2+} and \ce{C2H+} ions: The reported photoionization threshold of \ce{C2H2} is $11.4$ \textrm{eV} and appearance energy of \ce{C2H+} is $16.8$ \textrm{eV} \cite{Hayaishi_JPhysB_1982, Ono_JCP_1981} which can be associated with the \ce{C2H2+} states, $X^{2}\Pi_{u}$ and $A^{2}\Sigma_{g}^{+}$, respectively. For photoions whose appearance energies are higher than that of \ce{C2H2+} and \ce{C2H+}, $>16 \textrm{eV}$, their yield and photoionization efficiencies are also relatively lower (cf. table.\ref{table1}). We identify \ce{C2+} ion originating from two photodissociation channels:
	\begin{equation}\label{C2_diss_1}
		\ce{C2H2}+h\nu \rightarrow \ce{C2+} + \ce{H2} + e^{-}
	\end{equation}, and
	\begin{equation}\label{C2_diss_1}
		\ce{C2H2}+h\nu \rightarrow \ce{C2+} + 2\ce{H} + e^{-}
	\end{equation}
	with appearance energies of $18.1$ \textrm{eV} and $22.7$ \textrm{eV}, respectively\cite{Hayaishi_JPhysB_1982}. We infer these  mechanisms noting that the observed difference between these two \ce{C2+} appearance energies matches the dissociation energy of the \ce{H2} molecule\cite{Hayaishi_JPhysB_1982}. Similarly, for \ce{CH+} ion, there are two distinct appearance energies at $20.7$ \textrm{eV} and $24.1$ \textrm{eV} arising due to the following photoionization channels:
	\begin{equation}\label{CH_diss_1}
		\ce{C2H2}+h\nu \rightarrow \ce{CH+} + \ce{CH} + e^{-}
	\end{equation}, and
	\begin{equation}\label{CH_diss_1}
		\ce{C2H2}+h\nu \rightarrow \ce{CH+} + \ce{C} + \ce{H} + e^{-}
	\end{equation}
	, respectively\cite{Hayaishi_JPhysB_1982}. The appearance energies of the \ce{CH2+} $+$ \ce{C} and \ce{C+} $+$ \ce{CH2} photodissociation channels are $19.4$ \textrm{eV}\cite{Hayaishi_JPhysB_1982} and $24.0$ \textrm{eV}\cite{Cooper_CP_1988}, respectively. However, previous electron impact ionization study on \ce{C2H2} reported the appearance energy of \ce{C+} ion at a lower energy of $20.42$ \textrm{eV}\cite{Plessis_1986}. Table.\ref{table1} is a concise summary of the  measured relative ionization efficiencies ($\eta^{x}$) of these photoion compared with literature \cite{Hayaishi_JPhysB_1982}, corresponding to the aforementioned ionization channels, affirming the reliability of our measurements. We are now in a position to obtain insights into these processes taking advantage of the photoelectron imaging correlated to each of these photoions. This enables us to derive insights into state-selective photo-fragmentation dynamics.

\subsection{Photoelectron energy spectra and state-specific dissociation dynamics}

	To understand the mechanisms underlying the photoionization of \ce{C2H2}, here we present the photoelectron energy spectra (PES) and photoelectron angular distributions (PADs) of the photoelectrons correlated to all the ionic products of \ce{C2H2} photoionization. Then, PES and PADs in coincidence with each of the product ions are presented to investigate the ionization channels leading to these product ions. This allows us to compare our work with earlier reports, whereas the ionization channel specific investigation is a particular specialty of this work. We first discuss the \ce{C2H2} photoionization and the ionization channels that produce most abundant ions, \ce{C2H2+} and \ce{C2H+}. In the latter part, we discuss rest of the photodissociation channels.
	
	\begin{figure}[]
		\centering
		\includegraphics[width=0.5\textwidth]{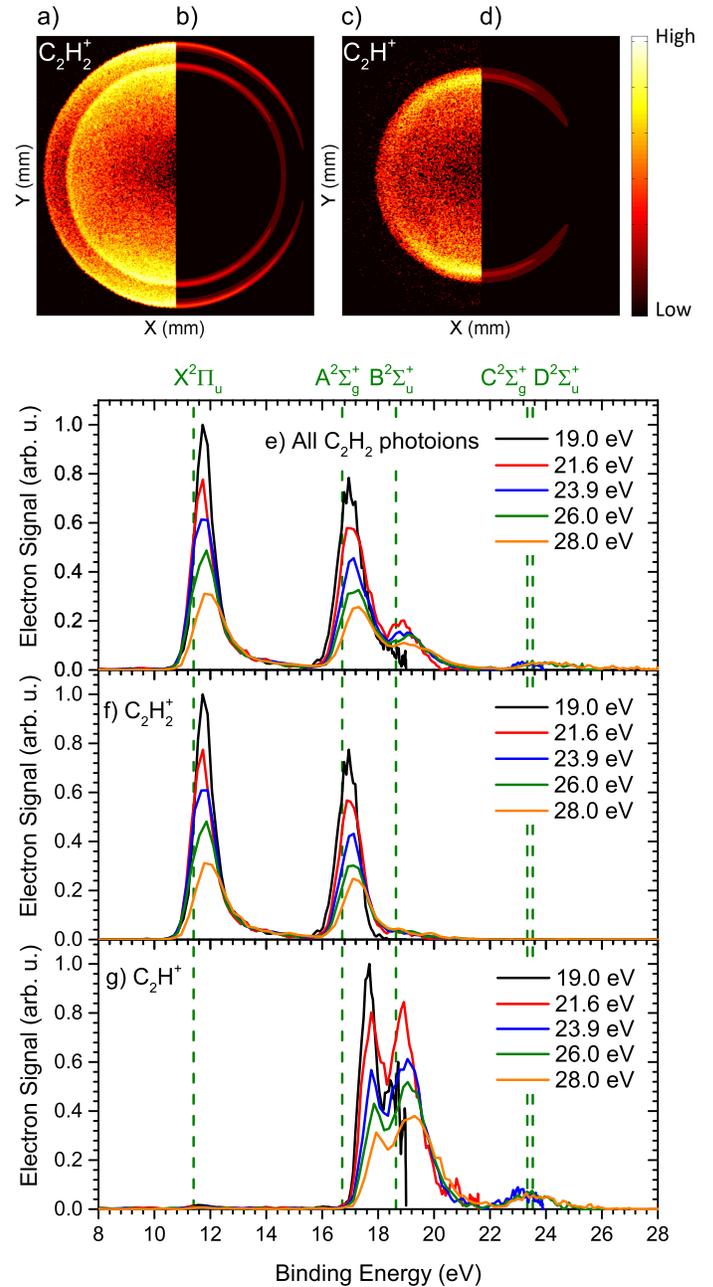}
		\caption{Photoelectron a) $\&$ c) VMI distribution, and b) $\&$ d) Abel-inverted distribution correlated to \ce{C2H2+} $\&$ \ce{C2H+} ions at $h\nu=28$ \textrm{eV}, in logarithmic color scale. Photoelectron energy spectra (PES): e) cumulative spectra of electrons summed over all photoions from \ce{C2H2} ionization, f) correlated specifically to the \ce{C2H2+} ion and g) the \ce{C2H+} ion, respectively. The vertical green dashed lines denote the ionization energies of \ce{C2H2+} states.}
		\label{Fig4_Ac}
	\end{figure}

    \begin{table*}[!htp]\centering
		\scriptsize
		\caption{State-specific binding energies (BE) correlated to all \ce{C2H2} photoions and in coincidence with specific ions at different photon energies ($h\nu$)}
		\label{table2}
		\begin{tabular}{l|rrrr|rrr|rrrr}\toprule
			\multirow{3}{*}{$h\nu$ (\textrm{eV})} &\multicolumn{10}{c}{BE (\textrm{eV})} \\\cmidrule{2-11}
			&\multicolumn{4}{c}{All \ce{C2H2} photoions} &\multicolumn{3}{c}{\ce{C2H2+}} &\multicolumn{3}{c}{\ce{C2H+}} \\\cmidrule{2-11}
			&$X^{2}\Pi_{u}$ &$A^{2}\Sigma_{g}^{+}$ &$B^{2}\Sigma_{u}^{+}$ &$C^{2}\Sigma_{g}^{+},D^{2}\Sigma_{u}^{+}$ &$X^{2}\Pi_{u}$ &$A^{2}\Sigma_{g}^{+}$ &$B^{2}\Sigma_{u}^{+}$ &$A^{2}\Sigma_{g}^{+}$ &$B^{2}\Sigma_{u}^{+}$ &$C^{2}\Sigma_{g}^{+},D^{2}\Sigma_{u}^{+}$ \\\midrule
			19.0 &11.79 $\pm$ 0.48 &16.95 $\pm$ 0.34 &17.49 $\pm$ 0.28 &\textemdash &11.79 $\pm$ 0.48 &16.92 $\pm$ 0.35 &\textemdash &17.62 $\pm$ 0.27 &18.54 $\pm$ 0.11 &\textemdash \\
            21.6 &11.72 $\pm$ 0.61 &17.09 $\pm$ 0.45 &18.76 $\pm$ 0.40 &\textemdash &11.72 $\pm$ 0.61 &17.01 $\pm$ 0.45 &18.46 $\pm$ 0.42 &17.67 $\pm$ 0.44 &18.86 $\pm$ 0.40 &\textemdash \\
            23.9 &11.78 $\pm$ 0.84 &17.14 $\pm$ 0.48 &17.49 $\pm$ 0.28 &23.29 $\pm$ 0.14 &11.78 $\pm$ 0.84 &17.09 $\pm$ 0.48 &18.96 $\pm$ 0.46 &17.72 $\pm$ 0.47 &19.00 $\pm$ 0.46 &23.08 $\pm$ 0.18 \\
            26.0 &11.83 $\pm$ 1.09 &17.16 $\pm$ 0.54 &17.49 $\pm$ 0.28 &23.92 $\pm$ 0.35 &11.83 $\pm$ 1.09 &17.12 $\pm$ 0.55 &18.87 $\pm$ 0.48 &17.78 $\pm$ 0.52 &19.10 $\pm$ 0.48 &23.44 $\pm$ 0.38 \\
            28.0 &12.01 $\pm$ 1.28 &17.27 $\pm$ 0.69 &18.76 $\pm$ 0.40 &24.24 $\pm$ 0.44 &12.01 $\pm$ 1.28 &17.22 $\pm$ 0.69 &18.98 $\pm$ 0.55 &17.88 $\pm$ 0.63 &19.23 $\pm$ 0.54 &23.60 $\pm$ 0.45 \\
			\bottomrule
		\end{tabular}
	\end{table*}
	
	\begin{table*}[h]\centering
		\scriptsize
		\caption{Comparison of equivalent state-selective ionization cross sections ($\sigma_{i}$) of different states ($i$) correlated to cumulative all \ch{C2H2} photoions with TDDFT calculation\cite{Fronzoni_CPL_2004}}
		\label{table3}
		\begin{tabular}{l|rr|rr|rr|rrr}\toprule
			\multirow{3}{*}{$h\nu$ (\textrm{eV})} &\multicolumn{8}{c}{$\sigma_{i}$ (\textrm{Mb})} \\\cmidrule{2-9}
			&\multicolumn{2}{c}{$i=X^{2}\Pi_{u}$} &\multicolumn{2}{c}{$i=A^{2}\Sigma_{g}^{+}$} &\multicolumn{2}{c}{$i=B^{2}\Sigma_{u}^{+}$} &\multicolumn{2}{c}{$i=C^{2}\Sigma_{g}^{+}, D^{2}\Sigma_{u}^{+}$} \\\cmidrule{2-9}
			&Current &TDDFT\cite{Fronzoni_CPL_2004} &Current &TDDFT\cite{Fronzoni_CPL_2004} &Current &TDDFT\cite{Fronzoni_CPL_2004} &Current &TDDFT\cite{Fronzoni_CPL_2004} \\\midrule
			19.0 &12.72 $\pm$ 0.32 &13.45 &8.94 $\pm$ 0.23 &15.57 &4.90 $\pm$ 0.12 &4.40 &\textemdash &\textemdash \\
            21.6 &12.36 $\pm$ 0.31 &12.12 &11.59 $\pm$ 0.29 &9.86 &5.09 $\pm$ 0.13 &5.50 &\textemdash &\textemdash \\
            23.9 &11.64 $\pm$ 0.29 &8.85 &9.08 $\pm$ 0.23 &8.23 &4.17 $\pm$ 0.10 &4.44 &0.49 $\pm$ 0.01 &0.89 \\
            26.0 &10.20 $\pm$ 0.26 &7.68 &7.09 $\pm$ 0.18 &7.02 &4.70 $\pm$ 0.12 &3.74 &1.03 $\pm$ 0.03 &1.15 \\
            28.0 &5.31 $\pm$ 0.13 &6.87 &4.79 $\pm$ 0.12 &6.04 &3.49 $\pm$ 0.09 &3.38 &1.38 $\pm$ 0.03 &1.18 \\
			\bottomrule
		\end{tabular}
	\end{table*}
	
    In fig.\ref{Fig4_Ac}, panels a) and c) show the photoelectron VMI distributions and panels b) and d) show the Abel-inverted distributions correlated to \ce{C2H2+} and \ce{C2H+} ions, respectively, at $28$ \textrm{eV} photon energy. Fig.\ref{Fig4_Ac} e) shows the cumulative PES summed over all the photoelectrons associated to all the photoions resulting from \ce{C2H2} photoionization. There are four distinct peak-structures in the PES centered at $11.8, 17.0, 19.0$ and $23.5$ \textrm{eV} representing different ionized states of \ce{C2H2}. The vertical green dashed lines in fig.\ref{Fig4_Ac} e) present the known ionization energies of first five cationic states $X^{2}\Pi_{u}$, $A^{2}\Sigma_{g}^{+}$, $B^{2}\Sigma_{u}^{+}$, $C^{2}\Sigma_{g}^{+}$ and $D^{2}\Sigma_{u}^{+}$ at $11.4$, $16.71$, $18.64$, $23.33$ and $23.53$ \textrm{eV}, respectively\cite{Wells_JCP_1999a}. Therefore, in this photon energy range ($19-28$ \textrm{eV}), these five states are mainly populated upon photoionization of \ce{C2H2}. Interestingly, PES correlated to the unfragmented \ce{C2H2^+} ion (cf. fig.\ref{Fig4_Ac} f)) do not have the fourth peak corresponding to the $C$ and $D$ states, evidencing that the unfragmented \ce{C2H2^+} ion is only produced from first three states. In contrast, PES correlated to \ce{C2H^+} ion (cf. fig.\ref{Fig4_Ac} g)) indicate that only the higher excited states, excluding $X$, lead to the \ce{C2H+} fragment, which in addition release a neutral \ce{H}. This observation reveals the mechanism underlying the previously reported appearance energy of \ce{C2H+} at $16.8$ \textrm{eV}\cite{Hayaishi_JPhysB_1982, Ono_JCP_1981}. Coincident photoelectron imaging in forthcoming discussions will reveal further details of the dynamics in this channel and others.
	
	In order to determine the binding energies (BE), multiple Gaussian functions are fitted to the PES; peak positions and relative intensities associated with different maxima  are determined. Owing to the finite energy resolution of the VMI spectrometer, we are not able to distinguish between the closely spaced $C$ and $D$ states. Therefore, we address the properties of this peak by labelling it as $C,D$. Table.\ref{table2} presents the binding energies corresponding to different \ce{C2H2+} states obtained from the fitting of PES correlated to cumulative all \ce{C2H2} photoions as well as spectra in coincidence with specific photoions, \ce{C2H2^+} and \ce{C2H^+}, respectively.
	
	In fig. \ref{Fig4_Ac} e)- g), we note upward shifts in PES peaks corresponding to the ionic state, $X$, compared to its ground vibrational level shown by the vertical green dashed line at $11.4$ \textrm{eV}. This upward shift in BE of the $X$ state can be attributed to the photoionization of \ce{C2H2} into the higher vibrational levels, $\nu_{2}=2$ and $3$, belonging to the ground ionized state ($X$) with  energies $11.85$ and $12.1$ \textrm{eV}, respectively\cite{Avaldi_JESRP_1995}. However, it should be noted that, the finite energy resolution of the spectrometer leads to significant widths in the reported binding energies, cf. table.\ref{table2}. From table.\ref{table2}, we see that the PES peaks in the spectra in coincidence with cumulative all \ce{C2H2} photoions, and those correlated to \ce{C2H2+} are nearly at the same positions for $X$ and $A$ states, whereas maxima in spectra correlated to $A$ and $B$ states in \ce{C2H+} are significantly shifted towards higher binding energies by $\sim0.66$ \textrm{eV} and $\sim0.23$ \textrm{eV}, respectively, as compared to the same in \ce{C2H2+}, also evident in panel g) of  fig. \ref{Fig4_Ac}. Thus, the additional binding energy is expended in climbing up the vibrational manifold of the $A$ and $B$ states of the \ce{C2H2^+} ion leading up to the dissociation into the \ce{C2H^+} ion and the neutral \ce{H}  \cite{Servais_CPL_1995}. For this dissociative ionization channel, the contributions of the higher excited states ($C,D$) are also significantly enhanced.
	
	\begin{figure}[t]
		\centering
		\includegraphics[width=0.5\textwidth]{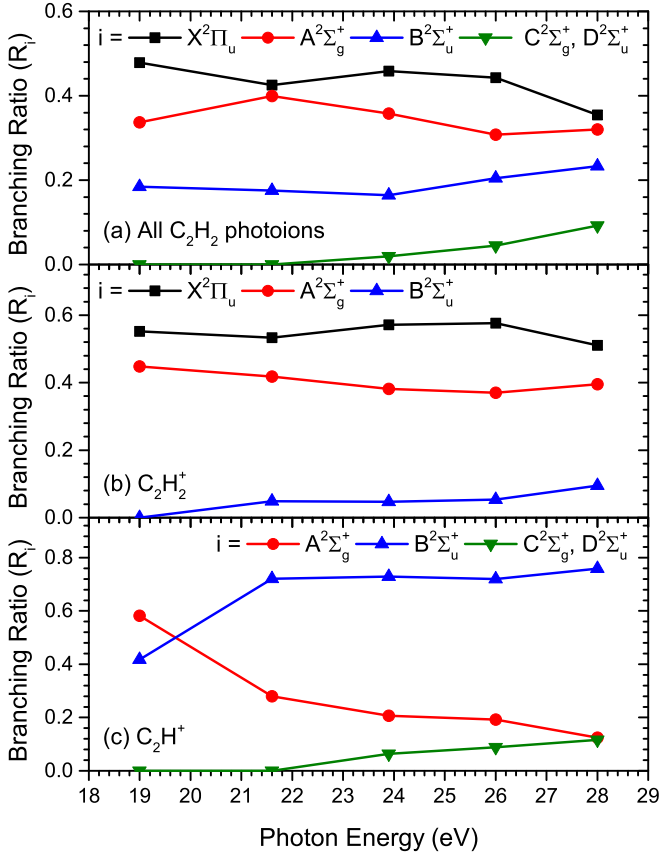}
		\caption{State-selective branching ratios ($R_{i}$) as a function of photon energy defined by equations \ref{eqn_F} and \ref{eqn_R_i} for different photoionization channels correlated to a) cumulative all \ce{C2H2} photoion, b) \ce{C2H2+} and c) \ce{C2H+} ions. The state-selective branching ratios, $R_{i}$,  are labelled by $i$ which correspond to the states, $X, A, B$ and $C,D$ of the primary photoion \ce{C2H2+} for different photoionization channels producing \ce{C2H2+}, \ce{C2H+}, \ce{C2+}, \ce{CH2+}, \ce{CH+} and \ce{C+} ions. The dependence of state-selective branching ratios ($R_{i}$) on photon energy reveal the role of possible resonances involved, as discussed in the text.}
		\label{Fig5_Ac}
	\end{figure}
	
	Furthermore, even though we are able to decipher the  electronic states resulting from \ce{C2H2} photoionization, particularly in the case of \ce{C2H2+} and \ce{C2H+} product ions, it is difficult to discern the  exact ionization process for the following reasons without additional knowledge: Both the direct photoionization and indirect autoionization processes lead to the same final electronic state; photoelectron kinetic energies emerging from the final state remain identical irrespective of the ionization mechanisms. However, the partial ionization cross sections of the final states and the associated photoelectron angular distributions will depend on the details of the ionization processes. At photon energies near the autoionization resonances, we may expect to see the impact of  resonances both in the partial ionization cross sections of these states and in the associated photoelectron angular distributions (PADs). Therefore, to discern the \ce{C2H2} photoionization processes, we will discuss photon energy dependent partial ionization cross sections in terms of state-selective branching ratios and photoelectron asymmetry parameter of these states in detail. To the best of our knowledge, state-selective branching ratios and  photoelectron asymmetries for different photoionization channels of \ce{C2H2} as a function of photon energy are reported for the first time.
	
	To determine the state-selective branching ratios ($R_{i}$) as a function of photon energy for different photoionization channels, we use the following method: First, the PES correlated to a specific photoionization channel for a given photon energy, $h\nu$, is fitted with the following multiple Gaussian functions, $F(E)$, of the form:
	\begin{equation}\label{eqn_F}
		F(E)=\sum_{i}C_{i}\frac{1}{\sqrt{2\pi}\sigma'_{i}}e^{-\frac{1}{2}(\frac{E-\textrm{BE}_{i}}{\sigma'_{i}})^{2}}
	\end{equation}
	where, $\textrm{BE}_{i}$, $\sigma'_{i}$ and $C_{i}$ represent the binding energy (BE), standard deviation and intensity of the $i^{th}$ state in the fitted PES, respectively.  Now, the branching ratios ($R_{i}$) of different ionic states ($i$) for the concerned photoionization channel at the photon energy, $h\nu$, can be written as:
	\begin{equation}\label{eqn_R_i}
    	R_{i} =\frac{C_{i}}{\sum_{i}C_{i}}
	\end{equation}
	These state-selective branching ratios ($R_{i}$) provide a normalized scaling for the relative intensities of individual ionic states ($i$) in a specific photoionization channel at a particular incident photon energy, where the total intensity summed over all the ionic states, $\sum_{i}R_{i}$, is equal to $1$.
	
	The state-selective branching ratios as a function of photon energy producing all \ce{C2H2} photoions are presented in fig.\ref{Fig5_Ac} a). As a trend, with increasing photon energy, the population of $X$ decreases while that of higher excited $C,D$ states increase. And, these ratios in the case of  $A$ and $B$ states, respectively, show  moderate photon energy dependence. These decreasing nature of relative intensities of $X$ and $A$ states can be attributed to the involved $2p$ atomic orbital. On the other hand, increasing state-selective branching ratios of $B$ and $C,D$ states are due to the involved $2s$ atomic orbital \cite{Holland_JESRP_1998}. In addition, a weak local maximum is observed in the $A$ state around $21.6$ \textrm{eV}. A comparison between the equivalent state-selective ionization cross sections ($\sigma_{i}$) of different states ($i$) obtained from our study with the same from theoretical time dependent density functional (TDDFT) calculation\cite{Fronzoni_CPL_2004} as a function of photon energy is reported in table.\ref{table3}. To calculate the equivalent state-selective ionization cross sections, we implement the following method: First, the state-selective relative ionization efficiencies ($\eta_{i}$) of different ionic states ($i$) for \ce{C2H2} photoionization at a particular photon energy are calculated as,
	\begin{equation}\label{eqn_eta_i}
	    \eta_{i}= \eta \times R_{i}
	\end{equation}
	where, $\eta$ and $R_{i}$ are the total relative ionization efficiency and the branching ratio of the corresponding ionic state ($i$). $\eta$ is related to cumulative efficiency of all photoions resulting from \ce{C2H2} ionization at the corresponding photon energy (cf. table.\ref{table1}). We take advantage of previously reported partial photoionization cross sections of \ce{C2H2} in the work of \citeauthor{Cooper_CP_1988}\cite{Cooper_CP_1988}, to determine the state-selective ionization cross sections, which are a fraction of the total cross section. Implementing this scheme, we calculated these state-selective ionization cross-sections ($\sigma_{i}$) from $\eta_{i}$ using the relation, $\sigma_{i} \; \textrm{(in Mb)} = (29.38 \pm 0.74)\times \eta_{i} \; \textrm{(in arb. u.)}$. The success of the measurements and this procedure is vindicated by the agreement of these experimental state-selective ionization cross sections with the theoretical calculation.
	
	To our advantage, the PEPICO technique enables the measurement of state-selective branching ratios associated with the photoionization channels producing \ce{C2H2+} and \ce{C2H+} ions as a function of photon energy, see fig.\ref{Fig5_Ac} b) and c). This leads to a comprehensive picture of ionization and dissociation in the photo-fragmentation process examining the PES correlated to \ce{C2H2+} and \ce{C2H+} ions. As observed from the PES (cf. fig.\ref{Fig4_Ac}) the higher excited ionic states, $(C,D)$, do not leave behind unfragmented \ce{C2H2+} ions, the lowest ionized state, $X$, does not participate in the dissociation process to produce \ce{C2H+}. For the \ce{C2H2+} ion, the state-selective branching ratio of $X$ state dominates over the same of $A$ and $B$ states. The state-selective branching ratios of the $X$ and $A$ states slightly decrease with increasing photon energy and are significantly higher than that of the $B$ state which slightly increases with increasing photon energy.  On the other hand, for \ce{C2H+} ion, the contributions of $B$ states are dominant over the $A$ and $C,D$ states. Beyond the photon energy of $\sim 21$ \textrm{eV}, the state-selective branching ratio of $A$ largely remain independent of photon energy. On the other hand, opposite behaviors are seen for $B$ and $C,D$ states, in which the branching ratio of $B$ decreases and the same of $C,D$ increases with increasing photon energy.
	
\subsection{Photoelectron angular distributions: fragment- and state-selected}
	
	\begin{figure}[t]
		\centering
		\includegraphics[width=0.5\textwidth]{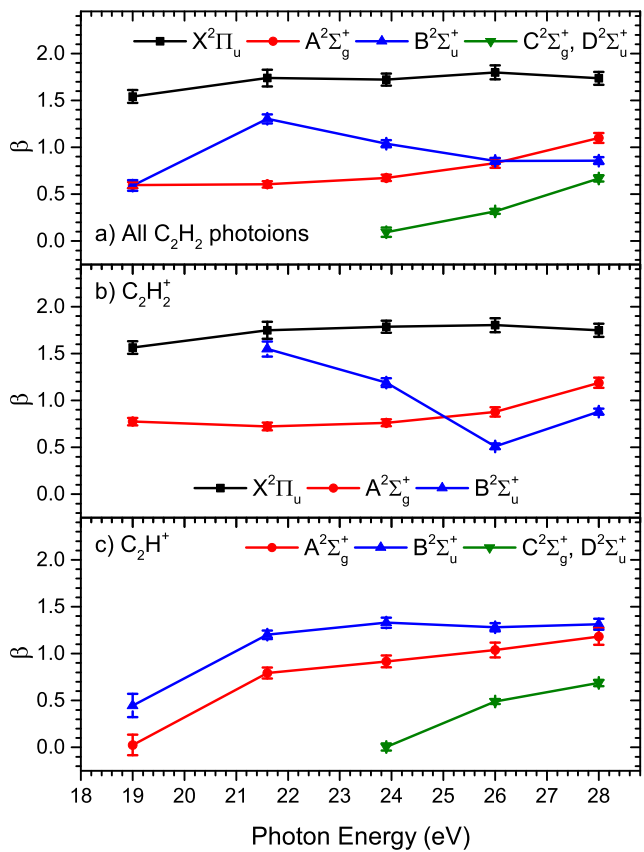}
		\caption{Photoelectron asymmetry parameter ($\beta$) of different states correlated to a) cumulative all \ce{C2H2} photoions b) \ce{C2H2^+} ion and c) \ce{C2H^+} ion, respectively, as a function of photon energy.}
		\label{Fig6_Ac}
	\end{figure}
	
	The most significant aspect of correlated photoelectron imaging is the opportunity it provides to examine photoion- and state-specific PADs. This immediately reveals the variations of the asymmetry parameters ($\beta$) of the photoelectron angular distributions correlated to different electronic states. Fig.\ref{Fig6_Ac} a), b) and c) depict $\beta$ as a function of photon energy; these are determined for the cases of photoelectrons in coincidence with cumulative all \ce{C2H2} photoions as well as PADs correlated to \ce{C2H2+} and \ce{C2H+} ions, respectively. Since the $\pi$-electron usually leads to a higher degree of asymmetry than that of $\sigma$ electron ejection\cite{Kreile_CPL_1980}, the observed large value of $\beta$ for $X$ state originating from the ionization of the HOMO ($1\pi_{u}$) is justified. Considering the low relative photoionization efficiencies of some of the channels, the VMI technique plays an important role in the measurement of $\beta$ parameters in which photoelectrons are collected over the entire solid angle. In this case, the measured $\beta$ ($\sim1.5$) for $X$ states is higher than the corresponding values obtained from the previous experimental studies where angle-resolved spectra were recorded using hemispherical electron analyzers \cite{Kreile_CPL_1980, Holland_JESRP_1998}.
	
	Two significant trends underscore the behaviour and physics of the dependence of the asymmetry parameter ($\beta$) as a function of photon energy: i) For the $X$ and $A$ states, the absence of autoionizing resonances in the chosen photon energy region, $19...28$ \textrm{eV}, underlies the observation of a weak dependence of the $\beta$, the asymmetry parameter on $h\nu$, cf. the black and red lines in fig.\ref{Fig6_Ac}. It is well known that autoionizing resonances influence PADs; there are no such channels decaying to the lower ionized states, $X$ and $A$, for $h\nu > 19$ \textrm{eV} \cite{levine_PRL_1983}, consistent with earlier results\cite{Holland_JESRP_1998, Lynch_JCP_1984}. ii) The higher ionized $B$ state shows considerable variations in the $\beta$ parameter with photon energy in the PADs of the cumulative all photoion distribution and those correlated with the \ce{C2H2+} ion, cf. the blue line in fig.\ref{Fig6_Ac} panels a) and b). We observe a local maximum around $21.6$ \textrm{eV} in the $\beta$ vs. $h\nu$ curve for photoelectrons in coincidence with cumulative all photoions and a minimum at $\sim26.0$ \textrm{eV} in the photon energy dependence of $\beta$ correlated to the \ce{C2H2+} ion. In previous works, a minimum was observed at $h\nu = 25$ \textrm{eV} for the $B$ state and reasoned as occurring due to interplay between $2\sigma_{u} \rightarrow k\sigma_{g}$ and $2\sigma_{u} \rightarrow k\pi_{g}$ transitions, where $k$ represents a state in the continuum \cite{Lynch_JCP_1984, Holland_JESRP_1998, Fronzoni_CPL_2004}. Therefore, it is surprising that we do not observe the minimum corresponding to the $B$ state curve for the dissociation product, \ce{C2H+} ion; there is no such structure in the $\beta$ vs. $h\nu$ curve. Rather, we only observe a nearly constant $\beta$  with increasing photon energy, cf. the blue line in fig.\ref{Fig6_Ac} c). This leads us to the conclusion that the observed local minimum around $26$ \textrm{eV} in the $\beta$ vs. $h\nu$ curve for \ce{C2H2+} ion is related  $2\sigma_{u}\rightarrow k\sigma_{g}, k\pi_{g}$ autoionizing resonances which play a prominent role in the formation of \ce{C2H2+} upon photoionization of the parent molecule. While \ce{C2H+} is formed by the dissociation of \ce{C2H2+}, the absence of this minimum (at $\sim26$ \textrm{eV}) in the $\beta$ associated with the former indicates formation \ce{C2H+} ion from non-autoionizing states. A fraction of the population of \ce{C2H2} participates in the aforementioned autoionizing resonance which is left undissociated when it decays, very likely, to lower vibrational states. However, \ce{C2H+} is formed by dissociation from a competing channel which proceeds through a population of the higher vibrational states of \ce{C2H2+} upon direct ionization. This observation motivates further theoretical investigations including multichannel interactions considering autoionizing states and nuclear dynamics. Finally, for the $C,D$ states, the value of $\beta$ increases steadily with increasing photon energy for cumulative all ions and \ce{C2H+} ions (green line in fig.\ref{Fig6_Ac} a), c)). Since these states do not produce unfragmented \ce{C2H2+}, autoionizing channels play no role.
	
	\begin{table*}[h]\centering
		\scriptsize
		\caption{Comparison of asymmetry parameters ($\beta$) correlated to the photoelectrons in coincidence with cumulative all \ce{C2H2} photoions with previous theoretical studies at different photon energies ($h\nu$)}
		\label{table4}
		\begin{tabular}{l|rrr|rrr|rrr|rrrr}\toprule
			\multirow{3}{*}{$h\nu$ (\textrm{eV})} &\multicolumn{12}{c}{$\beta$} \\\cmidrule{2-13}
			&\multicolumn{3}{c}{$X^{2}\Pi_{u}$} &\multicolumn{3}{c}{$A^{2}\Sigma_{g}^{+}$} &\multicolumn{3}{c}{$B^{2}\Sigma_{u}^{+}$} &\multicolumn{3}{c}{$C^{2}\Sigma_{g}^{+},D^{2}\Sigma_{u}^{+}$} \\\cmidrule{2-13}
			&Current &TDDFT\cite{Fronzoni_CPL_2004} &MCSCF\cite{Wells_JCP_1999a} &Current &TDDFT\cite{Fronzoni_CPL_2004} &MCSCF\cite{Wells_JCP_1999a} &Current &TDDFT\cite{Fronzoni_CPL_2004} &MCSCF\cite{Wells_JCP_1999a} &Current &TDDFT\cite{Fronzoni_CPL_2004} &MCSCF\cite{Wells_JCP_1999a} \\\midrule
			19.0 &1.54 &0.88 &0.74 &0.60 &0.31 &0.02 &0.59 &0.67 &0.59 &\textemdash &\textemdash &\textemdash \\
			21.6 &1.74 &1.03 &0.79 &0.61 &0.32 &0.15 &1.30 &1.16 &0.82 &\textemdash &\textemdash &\textemdash \\
			23.9 &1.72 &1.19 &1.06 &0.67 &0.36 &0.22 &1.04 &1.17 &1.22 &0.10 &1.06 &0.01 \\
			26.0 &1.80 &1.26 &\textemdash &0.83 &0.41 &0.30 &0.86 &1.13 &1.25 &0.32 &1.47 &-0.31 \\
			28.0 &1.74 &1.32 &\textemdash &1.10 &0.46 &0.41 &0.86 &1.08 &1.23 &0.67 &1.43 &-0.13 \\
			\bottomrule
		\end{tabular}
	\end{table*}
	
	Several theoretical studies performed hitherto\cite{Wells_JCP_1999a, Fronzoni_CPL_2004} addressed the variation of electronic state specific asymmetry parameter by considering multichannel interaction of electronic excitations in this photon energy range.  \citeauthor{Wells_JCP_1999a}\cite{Wells_JCP_1999a} implemented multichannel scattering methodology (MCSCF) where the partial cross sections and corresponding $\beta$ parameters of different ionic final states were calculated for different autoionization resonances. \citeauthor{Fronzoni_CPL_2004}\cite{Fronzoni_CPL_2004} used time dependent density functional method on a fixed nuclei geometry to calculated the autoionization channels and the asymmetry parameters. Table.\ref{table4} shows the comparison of the $\beta$ parameter obtained in our experiment with the previous theoretical studies \cite{Wells_JCP_1999a, Holland_JESRP_1998}. Only for the $B$ state, reasonable agreement between our experimental result and the theoretical calculation is observed. While for other ionized states our $\beta$ values are quite different from the same calculated from theory. It should be noted that both the theoretical studies estimate different values of $\beta$ for the $X, A$ and $C,D$ states, with reasonable agreement only for $B$ state.
	
\subsection{Higher-excited states: binding energies and dissociation channels}
	
	\begin{figure}[]
		\centering
		\includegraphics[width=0.5\textwidth]{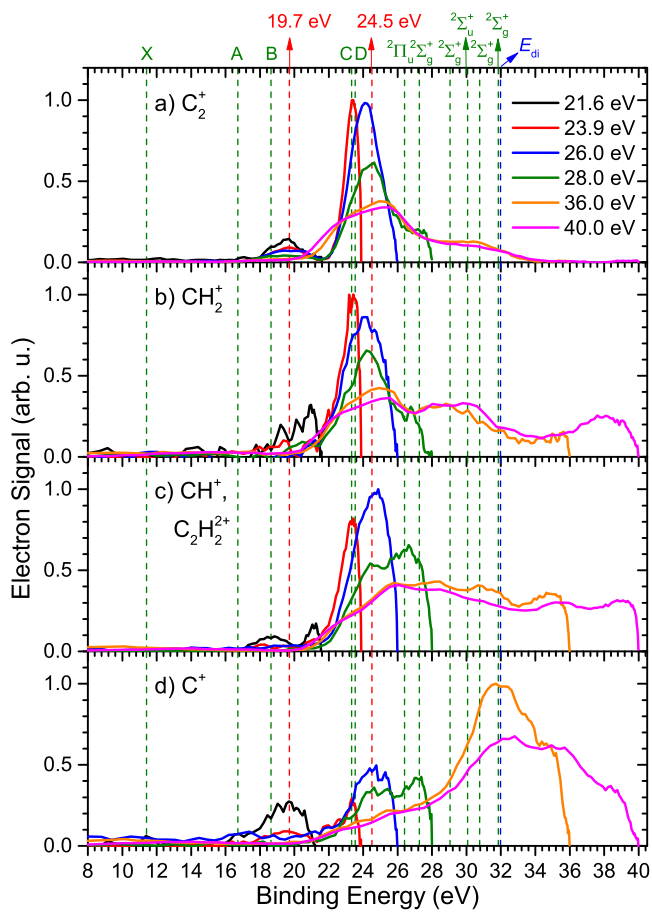}
		\caption{The photoelectron energy spectra correlated to a) \ce{C2+}, b) \ce{CH2+}, c) \ce{CH+}, \ce{C2H2^2+} and d) \ce{C+} ions, respectively, at different photon energies. The vertical green dashed lines show the binding energies of singly ionized states of \ce{C2H2}\cite{Wells_JCP_1999b}. Whereas the blue dashed line shows the double ionization energy ($E_{\textrm{di}}$) of acetylene. The red vertical dashed lines show the new peaks centred at $24.5$ and $19.7$ \textrm{eV}. The integral areas of the PES correlated to these ions at different photon energies are proportional to the relative ionization efficiencies ($\eta^{x}$) of these photoions ($x$) at the respective photon energy, shown in table.\ref{table1}.}
		\label{Fig7_Ac}
	\end{figure}
	
	Before concluding this article, we discuss the photoionization channels that produce low ion-yields at $M/q = 24, 14, 13$ and $12$ in the photoion ToF mass spectra. Here, both the single and double ionization regimes of \ce{C2H2} are covered in the photon energy range, $19...40$ \textrm{eV}, to access the higher excited electronic states of \ce{C2H2+} leading to these low-yield ions. However, the previous sections, only the single ionization pathways of \ce{C2H2} were discussed by presenting the PES and PADs correlated to cumulative all \ce{C2H2} photoions and the high yield ions, \ce{C2H2+} and \ce{C2H+}. PES corresponding to these photoions are plotted in fig.\ref{Fig7_Ac} as a function of binding energy (BE), where BE is calculated by subtracting the kinetic energy of the detected electron from the incident photon energy. Since we detect only one emitted electron both for single and double ionization events, a correct assignment of electronic states is only possible for the cationic states (BE $<$ $E_{\textrm{di}}$) which result from single ionization events. To assign dicationic states (BE $\ge$ $E_{\textrm{di}}$) relevant to double ionization, it would be necessary to take into account the total kinetic energy carried by both the emitted electrons. In fig.\ref{Fig7_Ac}, the vertical green dashed lines show cationic states leading up to the blue dashed line showing the double ionization energy ($E_{\textrm{di}} \sim 32$ \textrm{eV}) of \ce{C2H2}.
	
	\begin{table*}[t]\centering
		\scriptsize
		\caption{State-selective photoelectron asymmetry parameters ($\beta$) correlated to \ce{C2+}, \ce{CH2+} and \ce{CH+}, and \ce{C+} ions at different photon energies ($h\nu$)}
		\label{table5}
		\begin{tabular}{lr|rrrr|rrrr|rrrr|rrrr|rrrrr}\toprule
        \multicolumn{2}{c}{\multirow{2}{*}{\ce{C2H2^+} state}} &\multicolumn{20}{c}{$\beta$} \\\cmidrule{3-22}
        & &\multicolumn{4}{c}{$h\nu=40$ \textrm{eV}} &\multicolumn{4}{c}{$h\nu=36$ \textrm{eV}} &\multicolumn{4}{c}{$h\nu=28$ \textrm{eV}} &\multicolumn{4}{c}{$h\nu=26$ \textrm{eV}} &\multicolumn{4}{c}{$h\nu=23.9$ \textrm{eV}} \\\cmidrule{1-22}
        state &BI (\textrm{eV}) &\ce{C2+} &\ce{CH2+} &\ce{CH+} &\ce{C+} &\ce{C2+} &\ce{CH2+} &\ce{CH+} &\ce{C+} &\ce{C2+} &\ce{CH2+} &\ce{CH+} &\ce{C+} &\ce{C2+} &\ce{CH2+} &\ce{CH+} &\ce{C+} &\ce{C2+} &\ce{CH2+} &\ce{CH+} &\ce{C+} \\\midrule
        $C^{2}\Sigma_{g}^{+}$ &23.33 &1.33 &1.12 &1.15 &0.89 &1.14 &0.84 &1.08 &0.41 &0.70 &0.38 &0.63 &0.37 &0.33 &0.18 &0.27 &0.36 &0.25 &0.19 &0.02 &-0.03 \\
        $^{2}\Pi_{u}$ &26.40 &0.47 &0.63 &1.02 &0.70 &0.04 &0.09 &0.96 &0.60 &0.07 &-0.01 &0.13 &0.22 &\textemdash &\textemdash &\textemdash &\textemdash &\textemdash &\textemdash &\textemdash &\textemdash \\
        $^{2}\Sigma_{g}^{+}$ &27.27 &0.19 &0.13 &0.91 &1.12 &0.02 &0.04 &0.74 &0.68 &0.14 &-0.27 &0.03 &0.16 &\textemdash &\textemdash &\textemdash &\textemdash &\textemdash &\textemdash &\textemdash &\textemdash \\
        $^{2}\Sigma_{g}^{+}$ &29.04 &0.29 &1.09 &0.61 &0.83 &0.29 &0.20 &0.40 &0.36 &\textemdash &\textemdash &\textemdash &\textemdash &\textemdash &\textemdash &\textemdash &\textemdash &\textemdash &\textemdash &\textemdash &\textemdash \\
        $^{2}\Sigma_{u}^{+}$ &30.06 &0.44 &0.38 &0.55 &0.73 &0.34 &0.67 &0.43 &0.63 &\textemdash &\textemdash &\textemdash &\textemdash &\textemdash &\textemdash &\textemdash &\textemdash &\textemdash &\textemdash &\textemdash &\textemdash \\
        $^{2}\Sigma_{g}^{+}$ &30.77 &0.36 &0.08 &0.54 &0.82 &0.29 &0.03 &0.59 &0.43 &\textemdash &\textemdash &\textemdash &\textemdash &\textemdash &\textemdash &\textemdash &\textemdash &\textemdash &\textemdash &\textemdash &\textemdash \\
        $^{2}\Sigma_{g}^{+}$ &31.85 &0.25 &-0.08 &0.36 &0.60 &0.45 &0.17 &0.23 &0.36 &\textemdash &\textemdash &\textemdash &\textemdash &\textemdash &\textemdash &\textemdash &\textemdash &\textemdash &\textemdash &\textemdash &\textemdash \\
        24.5 eV &24.50 &0.67 &-0.33 &0.75 &0.84 &0.68 &0.21 &0.73 &0.60 &0.27 &0.79 &0.60 &1.06 &0.08 &0.19 &0.02 &0.11 &\textemdash &\textemdash &\textemdash &\textemdash \\
        \bottomrule
		\end{tabular}
	\end{table*}	
	
	Since, the studied photon energies cover the spectral range both below and above the $E_{\textrm{di}}$, we discuss these two regimes separately. For $h\nu<$ $E_{\textrm{di}}$, only single ionization of \ce{C2H2} molecule is possible. Therefore, photoion ToF mass peaks at $M/q = 24, 14, 13$ and $12$ correspond to \ce{C2+}, \ce{CH2+}, \ce{CH+} and \ce{C+} ions, respectively, which result from different fragmentation channels of \ce{C2H2+} ion. In fig.\ref{Fig7_Ac}, the PES correlated to all these ions have onset around the $B^{2}\Sigma_{u}^{+}$ which show intense peak structures around $24.5$ \textrm{eV} BE. This implies that the low-yield ions are produced from the higher excited states, $B$ and beyond, whereas the high-yield \ce{C2H2+} and \ce{C2H+} ions are found to be predominantly produced from the lower-lying $X, A$ and $B$ states. For $h\nu=21.6$ \textrm{eV}, we observe small PES peaks at BE around $B$ state. For \ce{C2+}, \ce{CH2+} and \ce{C+} ions, the corresponding PES peaks are centered at BE $=19.7$ \textrm{eV}, whereas for \ce{CH+} ion, the associated peak is coinciding with $B$ state (cf. fig.\ref{Fig7_Ac} c)). At $h\nu=23.9$ \textrm{eV}, PES correlated to all these photoions (red line in fig.\ref{Fig7_Ac}) have similar maxima around a binding energy of $23.3$ \textrm{eV} corresponding to the $C$ state. However, the associated photoelectron asymmetry parameters ($\beta$) for these ions are quite different (cf. table.\ref{table5}). The observed $\beta$ decreases from $0.25$ for \ce{C2+} ion to $-0.03$ for \ce{C+} ion. Similar peaks around $24.5$ \textrm{eV} are observed for photoionization at $26.0$ and $28.0$ \textrm{eV} (blue and green line in fig.\ref{Fig7_Ac}), along with two additional peak structures centered around $^{2}\Pi_{u}$ (BE = $26.40$ \textrm{eV}) and $^{2}\Sigma_{g}^+$ (BE = $27.27$ \textrm{eV}) states for $h\nu = 28.0$ \textrm{eV}. The peak around $24.5$ \textrm{eV} (vertical red dashed line in fig.\ref{Fig7_Ac}) cannot be associated with any reported cationic state, though. For \ce{C2+} and \ce{CH2+} ions, the $24.5$ \textrm{eV} peak dominates over the other two peaks assigned to $^{2}\Pi_{u}$ and $^{2}\Sigma_{g}^+$ states, while for \ce{CH+} and \ce{C+} ion all the three peaks are almost at equal intensity. For $^{2}\Pi_{u}$ (BE = $26.40$ \textrm{eV}) state, the observed photoelectron angular distributions are isotropic, which resulted in $\beta$ values close to zero. This is in contrast to the observed asymmetry ($\beta$) for $X^{2}\Pi_{u}$ state where we observe higher degree of asymmetry ($\beta\sim1.5$). However, the new peak around $24.5$ \textrm{eV} shows higher degree of asymmetry which increases from $0.27$ for \ce{C2+} to $1.06$ for \ce{C+} at $h\nu=28.0$ \textrm{eV}.
	
	For $h\nu>$ $E_{\textrm{di}}$, both single and double ionization of \ce{C2H2} are possible. Therefore, the mass peaks at $M/q = 24, 14, 13$ and $12$ correspond to fragment ions from both single and double ionization events. However, the contributions of single and double ionization events can be distinguished from the BE scale in the PES. For BE $<$ $E_{\textrm{di}}$, all the events are from single ionization processes which are discussed earlier. The relevant state-selective asymmetry parameters ($\beta$) are shown in table.\ref{table5} for $h\nu=36$ and $40$ \textrm{eV}. For BE $>$ $E_{\textrm{di}}$, the PES correspond to the detection of one of the two emitted electrons from double ionization of \ce{C2H2}. The assignment of electronic states is not feasible, as stated earlier. Fig.\ref{Fig7_Ac} a) shows the PES correlated to \ce{C2+} ion produced from the \ce{C2H2^2+} $\rightarrow$ \ce{C2^+} $+$ \ce{H+} $+$ \ce{H} dissociation process\cite{Thissen_JCP_1993}. Fig.\ref{Fig7_Ac} b) and d) show the PES corresponding to \ce{CH2^+} and \ce{C+} which result from the same photo-fragmentation channel \ce{C2H2^2+} $\rightarrow$ \ce{C+} $+$ \ce{CH2+}; this channel  involves the characteristic isomerization of acetylene \cite{Gaire_PRA_2014}. For \ce{C2H2^2+} and \ce{CH^+} photoions produced by non-dissociative and the dissociative (\ce{C2H2^2+} $\rightarrow$ \ce{CH+} $+$ \ce{CH+}) double ionization channels, respectively, the corresponding PES are shown in fig.\ref{Fig7_Ac} c) \cite{Gaire_PRA_2014}.
	
	For all the cationic states, we determine fragmentation channel specific photoelectron asymmetry parameters ($\beta$) from the angular distributions obtained by integrating photoelectron counts in Abel-inverted distributions considering a $1$ \textrm{eV} energy-window centered at the BE of each state. Table.\ref{table5} presents the details of the asymmetry parameter along with the BE of each state. This includes state-selective $\beta$ values which are distinct to photoionization channels with relatively low cross sections, producing
	\ce{C2+} , \ce{CH2+}, \ce{CH+} and \ce{C+} ions, contributing new knowledge about this important molecular system.
	
\section{Conclusion}
	
    Several intriguing dynamics of the acetylene-vinylidene system which play a central role in our understanding of proton migration and isomerization in the extreme ultraviolet, $19...40$ \textrm{eV} are uncovered. State-selective ionization pathways are identified for \ce{C2H2} photoionization. We observe that the unfragemented \ce{C2H2+} ion mainly results from the lower-lying $X, A$ and $B$ states, while photodissociation of \ce{C2H2+} from $A, B, C$ and $D$ states leads to \ce{C2H+} ion and neutral \ce{H}. Less abundant ions (\ce{C2+}, \ce{CH2+}, \ce{CH+} and \ce{C+}) are predominantly produced from even higher excited states, $B$ and beyond. For photoionization above the double ionization energy ($E_{\textrm{di}}$), these ions are produced due to fragmentation of \ce{C2H2^2+}. Below $E_{\textrm{di}}$, the isomerization of acetylene is addressed by presenting the PES and PAD in coincidence with \ce{CH^2+} ion. State-selective branching ratios and photoelectron asymmetry parameters ($\beta$) correlated to the relevant cationic states are reported as a function of photon energy for all the \ce{C2H2} ionization channels. Photon energy dependent photoelectron asymmetry parameter shows distinct patterns for the photoionizations leading to \ce{C2H2+} and \ce{C2H+} ions. Previously reported autoionizing resonance around $25$ \textrm{eV} decaying to $B$ state is found to be selective to the ionization pathway it proceeds. We observe this autoionization signature in the $\beta$ parameter only for unfragmented \ce{C2H2+} ion. Whereas photo-fragmentation channel producing \ce{C2H+} does not indicate such autoionization in the variation of its photoelectron angular distribution with photon energy. To understand the photoelectron dynamics in this system, particularly theoretical explorations combining both the nuclear and the electron dynamics are required. Finally, these results open avenues urging time-resolved studies of this important molecular system using table-top high-harmonic and free-electron laser pulses.

\section*{Conflicts of interest}
	Authors confirm that there are no conflicts of interest to declare.

\section*{Author contributions}

	VS, RG and SRK proposed and designed this research. SM, RG, HS, AD, RR, MC, MM, VS and SRK performed the experiment. SM, RG, RR, MM, VS and SRK contributed to analysis of the experimental data. SM, RG, RR, BB, AS, SS, MM, VS and SRK worked on the interpretation and phenomenology. SM, RG, RR, MC, SS, BB, MM, VS and SRK were involved in preparing the manuscript. BB, MM, RR, MC, VS and SRK contributed with scientific resources and funding towards the experimental realization and beamtime.
	
\section*{Acknowledgements}

	VS, RG and SRK are grateful to DST, India and ICTP, Trieste, for support (proposal \# 20165468) to carry out this campaign at  the Elettra Synchrotron facility. VS and SRK acknowledge financial support from the IMPRINT and DAE-BRNS scheme. SRK thanks the Max Planck Society for supporting this research via the Partner group. MM acknowledges support from the Carlsberg Foundation, and with SRK and VS for the funding from the SPARC programme, MHRD, India.
	
	\bibliography{MS_Mandal} 
	\bibliographystyle{rsc}
	
\end{document}